%% file: main.tex
\documentclass[runningheads]{llncs}

\include{macros}

\newacronym{CRT}{CRT}{Chinese Remainder Theorem}
\newacronym{HE}{HE}{Homomorphic Encryption}
\newacronym{LSB}{LSB}{Least Significant Bit}
\newacronym{MSB}{MSB}{Most Significant Bit}
\newacronym{SIMD}{SIMD}{Single Instruction Multiple Data}
\newacronym{RLWE}{R-LWE}{Ring Learning With Errors}
\newacronym{ML}{ML}{Machine Learning}
\newacronym{PPML}{PPML}{Privacy-Preserving Machine Learning}
\newacronym{SOTA}{SOTA}{State-Of-The-Art}

\begin{document}

\title{Generating One-Hot Maps under Encryption}

\authorrunning{Ehud Aharoni, Nir Drucker, Eyal Kushnir, 
Ramy Masalha, Hayim Shaul}
\titlerunning{Generating One-Hot Maps under Encryption}

\author{Ehud Aharoni\orcid{0000-0002-3647-1440} \and
Nir Drucker\orcid{0000-0002-7273-4797} \and
Eyal Kushnir\orcid{0000-0001-6123-0297} \and
Ramy Masalha\orcid{0000-0002-6808-5675} \and
Hayim Shaul\orcid{0000-0001-8432-0623}}

\institute{IBM Research - Israel} 

\maketitle

\begin{abstract}
One-hot maps are commonly used in the AI domain. Unsurprisingly, they can also bring great benefits to ML-based algorithms such as decision trees that run under Homomorphic Encryption (HE), specifically CKKS. Prior studies in this domain used these maps but assumed that the client encrypts them. Here, we consider different tradeoffs that may affect the client's decision on how to pack and store these maps. We suggest several conversion algorithms when working with encrypted data and report their costs. Our goal is to equip the ML over HE designer with the data it needs for implementing encrypted one-hot maps.
\end{abstract}


\section{Introduction}

Complying with regulations such as GDPR \cite{GDPR} and HIPAA \cite{HIPAA} can prevent organizations from porting sensitive data to the cloud. To this end, some recent \gls{PPML} solutions use \gls{HE}, which enables computing on encrypted data. The potential of HE can be observed in Gartner's report \cite{gartner}, which states that  50\% of large enterprises are expected to adopt HE by 2025 and also in the large list of enterprises and academic institutions that are actively engaging in initiatives like HEBench \cite{hebench} and \gls{HE} standardization efforts \cite{standard}. 

We illustrate the landscape of HE-based solutions by first describing one commonly used threat model. For brevity, we restrict ourselves to a basic scenario, which we describe next, but stress that our study can be used almost without changes in many other constructions. We consider a scenario that involves two entities: a user and a semi-honest cloud server that performs \gls{ML} computation on HE-encrypted data. The user can train a model locally, encrypt it, and upload it to the cloud. Here, the model architecture and its weights are not considered a secret from the user only from the cloud. Alternatively, the user can ask the cloud to train a model on his behalf over encrypted/unencrypted data and at a later stage, perform inference operations, again, on his behalf using the trained model. We note that in some scenarios, the model is a secret and should not be revealed to the user. In that case, only the classification or prediction output should be revealed. We also assume that communications between all entities are encrypted using a secure network protocol such as TLS 1.3 \cite{rfc8446}, i.e., a protocol that provides confidentiality, integrity, and allows the users to authenticate the cloud server. 

\gls{HE}-based solutions showed great potential in past research but they also come with some downsides. Particularly, they involve large latency costs that may prevent their vast adoption. These latency costs can sometimes be traded with other costs such as memory or bandwidth, where these trade-offs allow the users to find the right balance for them. Our study aims to extend the \gls{HE} toolbox with new utilities and trade-offs that will get \gls{SIMD}-based \gls{HE}-solutions one step further in being practical. 

\paragraph{Branching.} In general, a branching operation that navigates a program to a specific path is unsupported under \gls{HE}. Instead, algorithms such as decision trees that select specific branches based on comparison operations are required to select and go through all branches when executed under HE. They often do it by using comparisons and indicator masks. This increases the number of executed operations, which in decision trees increases the number of comparison operations that are costly when executed under HE. In Section \ref{sec:app}, we discuss some possible comparators and conclude, as prior research did, that the most efficient comparators are binary comparators, which only compare one bit of data at a time. This led many implementations over \gls{HE} to prefer using \textit{one-hot} encoding for the input data. 

\begin{definition}
A one-hot map that represents $n$ categories, is an $n$-bits vector $\vect{o} \in \{0,1\}^n$ where $\sum_{i=0}^{n-1}{\vect{o}[i]}=1$, i.e., all bits but one are 0. 
\end{definition}

Most prior arts e.g., \cite{usingonehot1,usingonehot2,usingonehot3,usingonehot4, usingonehot5, crf} assumed that these maps are precomputed by the user before encryption and thus their costs are negligible. However, in practice, this computation can lead to huge overhead costs of memory and bandwidth, which many applications try to avoid. Furthermore, in some applications, the data was encrypted and uploaded to the cloud way before the \gls{ML} processes are required, e.g., when collecting data from IoT devices. In this case, some pre-processing methods should be applied to convert the encrypted input from one format to another. Our goal is therefore to explore the different trade-offs when using different conversion methods and use cases.

\paragraph{Our contribution.} We explore trade-offs that are related to one-hot maps.
\begin{itemize}
    \item We compare different input representations and explain the benefit of every representation as well as when it should be used.
    \item We propose a new input representation that is based on the \gls{CRT}, which offers a new trade-off between latency and bandwidth.
    \item We propose several new conversion algorithms between different input representations.
    \item We implemented and experimented with our proposed methods. In addition, our code is available online \cite{helayers}.
\end{itemize}

\paragraph{Organization.}
The document is organized as follows. Section \ref{sec:bg} describes some preliminaries such as balanced trees and \gls{HE}. We describe possible input representations of data and our novel CRT representation in Section \ref{sec:input} and we describe our conversion methods in Section \ref{sec:conv}. Some applications that can use one-hot maps are detailed in Section \ref{sec:app}. Sections \ref{sec:input}-\ref{sec:app} present the methods we used regardless of the final packing in ciphertexts. To this end, Section \ref{sec:pack} describes tile tensors and some possible packing methods. Section \ref{sec:exp} describes the setup and results of our experiments and Section \ref{sec:conc} concludes the paper.

\section{Preliminaries and notation}\label{sec:bg}
We refer to numbers using low-case letters e.g., $a,b,c$, and we denote vectors by bold face letters e.g., $\vect{v}, \vect{w}$. To access a specific element $0 \le i < n$ in a vector $\vect{v}$ we use square brackets $\vect{v}[i]$. A dot-product operation of two vectors $\vect{v}_1,\vect{v}_2$ is denoted by $\angles{\vect{v}_1}{\vect{v}_2}$. One-hot maps are often represented using the letter $\vect{o}$ and greater maps (defined below) using the letter $\vect{g}$. A mathematical ring with addition $(+)$ and multiplication $(*)$ operations is denoted by $\R(+,*)$. In this paper, $\log$ refers to $\log_2$.

\subsection{Balanced binary trees}
Some of our algorithms use a balanced binary tree structure. We define the tree relations in the trivial way. For example, in the tree below, Root is the parent of Node1 and Node2, which are the sons of Root. Two nodes are siblings if they share the same first-order parent and we access them using the function \sibling, for example, $Node2 = Node1.\sibling()$ and $Root.\sibling() = \varnothing$. A path to Leaf1 includes all nodes from Leaf1 to Root, e.g., $Leaf1.\leafpath = \{Leaf1, Node1, Root\}$. The leaves are numbered from left to right and for a tree $T$ we access its $i$th leaf from the left using the function $T.\leaf[i]$. To get all nodes at a specific level we use the method \getNodesAtLevel, where the level indexing starts from 0. For example,
\begin{align}
& \getNodesAtLevel(0)=\{Root\} \\ 
& \getNodesAtLevel(1)=\{Node1, Node2\} \\
& \getNodesAtLevel(2)=\{Leaf1, Leaf2, Leaf3, Leaf4\}
\end{align}
Finally, to get the right or left sons of a node we use the methods \leftn and \rightn, respectively. We use the method $\val()$ to access the value of a node.

\tikzset{every tree node/.style={minimum width=2em,draw,blank},
         blank/.style={draw=none},
         edge from parent/.style=
         {draw,edge from parent path={(\tikzparentnode) -- (\tikzchildnode)}},
         level distance=1cm,
         sibling distance=5pt}

\begin{center}
\begin{tikzpicture}
\Tree [.Root
    [ Leaf1 Leaf2 ].Node1 
    [ Leaf3 Leaf4 ].Node2 
      ]
\end{tikzpicture}
\end{center}

\subsection{\acrfull{HE}}
\gls{HE} is a public-key encryption scheme that in addition to the usual functions $\Enc, \Dec$ (see details below) also provides functions to perform operations  on encrypted data (usually addition and multiplication), see survey in \cite{Halevi2017}.
The encryption operation $\Enc:\R_1 \rightarrow \R_2$ encrypts input plaintext from the ring $\R_1(+, *)$ into ciphertexts in the ring $\R_2(\oplus, \odot)$ and its associated decryption operation is $\Dec:\R_2 \rightarrow \R_1$. Informally, an \gls{HE} scheme is correct if for every valid input $x,y \in \R_1$: $\Dec(\Enc(x)) = x$, $\Dec(\Enc(x) \oplus \Enc(y)) = x + y$, and $\Dec(\Enc(x) \odot \Enc(y))  = x * y$, and is approximately correct (as in CKKS) if for some small $\epsilon > 0$ that is determined by the key, it follows that $|x - \Dec(\Enc(x))| \le \epsilon$. The addition and multiplication equations are modified in the same way. Note that the correctness equations should hold for every finite number of multiplications and additions. This paper, focuses on CKKS.

In \cite{alonSIMD} the authors showed how \gls{SIMD} operations can be implemented in lattice based cryptography. Later, similar ideas were implemented in Modern \gls{HE} instantiations such as BGV \cite{bgv}, B/FV \cite{bfv1, bfv2}, and CKKS~\cite{ckks2017} that rely on the complexity of the Ring-LWE problem \cite{rlwe} for security.
For schemes that support \gls{SIMD}, addition and multiplication are applied slot-wise on vectors.

\paragraph{HE Packing.} Some \gls{HE} schemes, such as CKKS \cite{ckks-rns}, operate on ciphertexts in a homomorphic \gls{SIMD} fashion. This means that a single ciphertext encrypts a fixed-size vector, and the homomorphic operations on the ciphertext are performed slot-wise on the elements of the plaintext vector. To utilize the SIMD feature, we need to pack and encrypt more than one input element in every ciphertext. The packing method can dramatically affect the latency (i.e., time to perform computation), throughput (i.e., number of computations performed in a unit of time), communication costs, and memory requirements. We further discuss packing one-hot vectors under \gls{HE} in Section \ref{sec:pack}.

\paragraph{Comparison under HE.}
Comparing numbers under \gls{HE} and specifically, CKKS (e.g., \cite{cheon2020efficient, lee2021minimax}) often relies on polynomial approximations of the $\Step()$ or $\Sign()$ functions, whereas the accuracy and performance of these methods rely on the degrees of these polynomials. We denote these comparison functions by $\EQ(x,y) = 1 \Leftrightarrow x = y$, and $\EQ=0$ otherwise. Note that other methods exists for schemes such as BGV and BFV such as \cite{bgvbfvcomparison, homcompare}.

\section{Input representations}\label{sec:input}
This section discusses different input representations for algorithms that may use one-hot maps. In general, the one-hot map slots can represent any set of categories (\Cat) e.g., $\Cat=['dog', 'cat', 'bird']$ or $\Cat=[0.9, 0.7, 1.57]$. To simplify our algorithms we denote by $[n]$ the set $\{0,1,\ldots,n-1\}$ and associate every element of \Cat to an element in $[n]$. One commonly used example for \Cat is the quantization of a floating point element in some fixed range. For example, a floating point value $a \in [1,2]$ can be converted to one of 6 categories: $\Cat_{a}=[1,1.2,1.4,1.6,1.8,2]$, where we can use the bijective function $\phi : \Cat_{a} \longrightarrow [n]$, where $\phi(a) = 5a - 4$. This map can easily be implemented under \gls{HE} if needed using one multiplication by a constant and one addition.

We consider several methods for achieving one-hot maps from categorical variables over $[n]$. For these, we consider six input representations: 1) One-hot map representation; 2) Numeric representation; 3) \gls{CRT} representation; 4) Hierarchical \gls{CRT} representation; 5) Numeric (Hierarchical) \gls{CRT} representation; 6) Binary representation.

\paragraph{One-hot map representation.} This is the simplest representation, where the client directly encodes every one of its number elements as a one-hot map, i.e., $n$ values are encrypted for each element. The issue with this method is that it is wasteful in bandwidth, assuming that $n$ is of medium size e.g., $n = 100$ will cause the client to encode $99$ zeros and one $1$ to represent just one element. When the network transport and storage costs are large, this method becomes unfavorable. One-hot maps were used for example in \cite{usingonehot1,usingonehot2,usingonehot3,usingonehot4, usingonehot5, crf}.
    
\paragraph{Numeric representation.} Here the client submits elements from $[n]$ using their integer representation. Here, the bandwidth and storage costs are low, but converting this representation to a one-hot map representation while encrypted on the cloud side is costly in terms of latency. 
It either requires $n$ costly \EQ operations or running some optimization such as polynomial interpolation (Section \ref{sec:conv}) that has multiplication-depth of at least $O(\log{n})$ and requires at least $n$ multiplications. This cost is high for large values of $n$.
    
\paragraph{CRT representation.} We propose a technique that offers a tradeoff between the above two approaches. It possesses medium bandwidth costs and medium latency costs. The idea is to submit one-hot maps of the \gls{CRT} representation of every element and expand them in an online phase by the server. The size of all the maps is much smaller compared to the size of a full-size one-hot map, and the latency to expand them is lower compared to transforming an integer element into a one-hot map. Specifically, for $a \in [n]$, where $n$ is a composite number with pairwise coprime factors, i.e., $n=n_1 \cdot n_2 \cdots n_k$, where $GCD(n_i, n_j)=1$ for $i \neq j$. In that case, we use a set of $k$ one-hot maps $\vect{o}_{a,i}$ of $n_i$ slots, respectively, where $\vect{o}_{a,i}[a \pmod{n_i}]=1$ and the other slots of $\vect{o}_{a,i}$ are set to 0. The total size of these maps is $\sum_{i=1}^{k}{n_i}$ which is much smaller than $n$. 

In practice, we can choose any $n$ even without coprime factors. To use the above method we embed the $[n]$-map in an $[m]$-map, where $n < m$ and $m$ has coprime factors. The extra slots are set to zero. For example, let $n=10{,}000$, we can choose to work with $m=2 \cdot 5\cdot 7\cdot 11\cdot 13=10{,}010$ and the number of required slots is $r=38$. There are different heuristics to find the smallest $m$ or $r$ values, one such can be to consider a brute-force search over all different combinations of the $\ell$ smallest primes, or to set a limit $t$ and consider all numbers $m \in [n,n+t]$.

\paragraph{Hierarchical \gls{CRT} representation.} The \gls{CRT} representation is limited to coprime factors. Here, we suggest another approach that uses \gls{CRT} representation in a hierarchical way, where we consider the coprime restriction separately at every hierarchy. Start from a number $n$, compute $n_1 = \ceil{\sqrt{n}}, n_2 = \ceil{\sqrt{n}}+1$.  Clearly, $n_1$ and $n_2$ are coprime. Recursively repeat the split process until reaching the ``desired'' hierarchy according to latency and bandwidth parameters. The client then encodes its value $a$ using the last layers' values and the cloud constructs the one-hot map in an iterative process. Example \ref{ex:hier} in Appendix \ref{sec:exhier} demonstrates this representation.

\paragraph{Numeric (hierarchical) \gls{CRT} representation.} In Example \ref{ex:hier} the client encrypts the 8 maps $\vect{o}_{a, *}$, another option is to encrypt their $8$ representative numerical values, e.g., $a_{*}$, in that case, only 8 elements are being sent. We call this representation the Numeric \gls{CRT} representation.

\paragraph{Binary representation.} The last representation uses binary decomposition instead of \gls{CRT}-based decomposition. Here, the client decomposes the number $a$ to its binary decomposition $a=\sum_{i=0}^{\log{(n)}-1}\vect{a}[i] \cdot 2^i$, the number of elements being sent is therefore $\log{(n)}$.

\section{Conversion methods.}\label{sec:conv}

Some applications assume that the input representation is static and choose the one that most fits their needs. In contrast, some applications assume that the data was uploaded in one format and needs to be translated into another. Here, we describe some conversion methods between input representations. Another reason to move between representations is when an application uses the data in different places, where at every place it prefers the data in a different format. Note that moving from any representation into the \gls{CRT} representation is hard as the modulo operation is slow and sometimes even unsupported under \gls{HE}.

\subsection{From one-hot to numeric representation}\label{sec:fromcrt}
Moving from a one-hot representation of a vector $\vect{o}_a$ that represents the classes $\Cat_{a}$ to its numeric representation $a$ can be done by simply computing $\angles{\Cat_{a}}{\vect{o}_a}$, which requires $n$ multiplications and a multiplication-depth of 1. In that case, there is no need to apply the $\phi$ transformation to $[n]$ before and after the operation. When $c_a$ is not encrypted, only $n$ cleartext-ciphertext multiplications are performed with a multiplication depth of 0.

\subsection{From CRT representation to one-hot maps}\label{sec:tocrt}
We now describe how to convert data in a CRT representation to one-hot maps. The input is the $k$ one-hot sub-maps $\vect{o}_{a,i}$, which we first duplicate $\frac{n}{n_i}$ times, respectively, to the duplicated vectors $(\vect{o}^d_{a,i})_{i\le k}$. Subsequently, we multiply all the duplicated maps to construct the final map by
$
\vect{o}_a[i] = \prod_{j=1}^{k}{\vect{o}^d_{a,j}[i]}
$
This product is often done by using a multiplication tree, where the cost is only $n\cdot k$ multiplications and a multiplication depth of $\log{k}$. 

\begin{example} 
Let $n=2\cdot 3 \cdot 5=30$ where $n_1=2$, $n_2=3$, $n_3=5$ then the one-hot representation of the number $17$ is 
\[
(0,0,0,0,0,0,0,0,0,0,0,0,0,0,0,0,1,0,0,0,0,0,0,0,0,0,0,0,0,0)
\]
and the associated \gls{CRT} maps are: 
\begin{align*}
\vect{o}_{a,1}=(0,1) & & \vect{o}_{a,2}=(0,0,1) & & \vect{o}_{a,3}=(0,0,1,0,0),
\end{align*}
which contains only $\sum_i{n_i}=2+3+5=10$ elements. To construct the full map from the \gls{CRT} maps we duplicate the entries of $\vect{o}_{a,1}$, $\vect{o}_{a,2}$, and $\vect{o}_{a,3}$, $30/2=15$, $30/3=10$, and $30/5=6$ times, respectively as follows
\begin{align*}
& \vect{o}^d_{a,1} = (0,1,0,1,0,1,0,1,0,1,0,1,0,1,0,1,0,\mathbf{1},0,1,0,1,0,1,0,1,0,1,0,1) \\
& \vect{o}^d_{a,2} = (0,0,1,0,0,1,0,0,1,0,0,1,0,0,1,0,0,\mathbf{1},0,0,1,0,0,1,0,0,1,0,0,1) \\ 
& \vect{o}^d_{a,3} = (0,0,1,0,0,0,0,1,0,0,0,0,1,0,0,0,0,\mathbf{1},0,0,0,0,1,0,0,0,0,1,0,0) 
\end{align*}
Finally, we compute 
\[
\vect{o}_{a} = (0,0,0,0,0,0,0,0,0,0,0,0,0,0,0,0,0,\mathbf{1},0,0,0,0,0,0,0,0,0,0,0,0)
\]
as expected.
\end{example} 

\paragraph{From Hierarchical \gls{CRT} to one-hot maps.} Unsurprisingly, converting an input from a Hierarchical \gls{CRT} representation to the one-hot maps representation uses recursively the \gls{CRT} method at every hierarchy. The hierarchical method has an advantage for large $n$ values.

\subsection{From numeric representation to one-hot maps}\label{sec:trees}
We present several methods for moving between a numeric representation to a one-hot map representation. The na\"ive approach relies on the \EQ operator, by checking whether an input $x$ equals one of the classes in \Cat, i.e., for every $i<n$ we compute $\vect{o}[i] = \EQ(\Cat[i], x)$. Assuming that $EqM$, $EqD$ be the number of multiplications and the multiplication depth of the $Eq$ operator, then such an algorithm requires  $EqM \cdot n$  multiplications with an overall multiplication depth of $EqD$. We suggest several alternatives that rely on Lagrange polynomial interpolations to transform a categorical element $a \in [n]$ to its one-hot map representation. Specifically, let $\vec{S}[c]=\prod_{i \in [n], i \neq c}{(c - i)}$, then the $c$'th bit of the one-hot map can be computed using the polynomial
\[
P_c(a)=\vec{S}[c]^{-1}\prod\limits_{i \in [n], i \neq c}{(a - i)},
\]
where $P_c(a)=1$ if $a=c$ and $P_c(a)=0$ otherwise.
Note that the denominator $\vec{S}[c]$ can be efficiently pre-computed because all of its inputs are public and known in advance. 

\paragraph{Efficiently compute the $n$ polynomials $P_c(a)$.} Every polynomial requires $n-2$ multiplications and $log{(n-2)}$ multiplication depth. The trivial way is to compute each polynomial separately, which requires $n(n-2)$ multiplications and $\log{(n-2)}$ multiplication-depth. However, note that every two polynomials share $n-3$ out of the $n-2$ terms. Thus, we propose other dedicated algorithms to perform the task more efficiently. Specifically, we suggest two tree-based algorithms.
For simplicity, we describe the case where $n$ is a power of two and we claim that it is possible to consider other options as well by adding dummy multiplications by one. Table \ref{tab:bounds} in Appendix \ref{sec:sc} shows the minimal and maximal values of a tree of $n$ nodes. The values are not too extreme when $n\le 2^5$.

\vspace{-10pt}
\subsubsection{Tree-based alternative 1.}

Our first alternative includes a tree-based algorithm, which we describe in Algorithm \ref{alg:tree1}. This is a ``shallow but big'' algorithm that requires $n \log{n}$ multiplications and having $\log{n}$ multiplication depth. The algorithm receives an input $x$ and an uninitialized balanced tree $T$ of depth $\log{n}$. The algorithm starts by initializing all leaves of the tree (Lines 2-3), then it initializes the rest of the tree (Lines 4-6), and finally, it computes the one-hot map using the pre-computed values. Figure \ref{fig:alg1init} illustrates the initialization process. Note that because in Line 9 the \leafpath method returns the nodes ordered from leaf to root, the multiplication depth of the entire algorithm is only $\log{n}$. While the same computation would have held if the path list was ordered from root to leaf, the overall multiplication depth would have resulted to be $2\log{n}$.

\begin{algorithm}[ht!]
    \caption{Shallow but big}
    \label{alg:tree1}
    \begin{algorithmic}[1]
        \Statex \textbf{Input:} $x$ the input ciphertext, $T$ a balanced binary tree of depth $\log{n}$
        \Statex \textbf{Output:} $o$ a one-hot map that represents $x$.
        \Statex \textbf{Assume:} $\vect{S}$, where $\vect{S}[c]=\prod_{i \in [n], i \neq c}{(c - i)}$
        \Procedure{Opt 1}{}
            \For{$c$ in $[n]$}
                \State $T.\leaf(i) = x-c$
            \EndFor
            \For{$l$ in $\log{n}-1$ to 0}
                \For{$v$ in $T.\getNodesAtLevel(l)$}
                    \State $v.val = (v.\leftn.\val) \cdot (v.\rightn.\val)$
                \EndFor
            \EndFor
            \For{$c$ in $0$ to $n-1$}
                \State $o[c] = \vect{S}[c]$
                \For{$v$ in $T.\leaf(c).\leafpath$}
                    \State $o[c] = o[c] * v.\sibling.\val$
                \EndFor
            \EndFor
            \State \Return $o$
        \EndProcedure
    \end{algorithmic}
\end{algorithm}

\vspace{-10pt}
\subsubsection{Tree-based alternative 2.}
\begin{figure}[ht!]
    \centering
     \begin{subfigure}[b]{0.48\textwidth}
        \centering
            \begin{tikzpicture}
            \Tree [.Root
                [ $(x-1)$ $(x-2)$ ].Node1 
                [ $(x-3)$ $(x-4)$ ].Node2 
                  ]
            \end{tikzpicture}
        \caption{Leaf initialization (Lines 2-3).}
    \end{subfigure}
     \begin{subfigure}[b]{0.48\textwidth}
        \centering
            \begin{tikzpicture}
                \Tree [.$(x-1)(x-2)(x-3)(x-4)$
                    [ $(x-1)$ \node[text=red]{$(x-2)$}; ].$(x-1)(x-2)$
                    [ $(x-3)$ $(x-4)$ ].\node[text=red]{$(x-3)(x-4)$};
                      ]
            \end{tikzpicture}
        \caption{Tree initialization (Lines 4-6).}
    \end{subfigure}
    \caption{Algorithm \ref{alg:tree1} initialization phase where $\Cat=[4]$ and $n=4$. Red nodes show the multiplications done to compute $o[1]$ on Line 11.}
    \label{fig:alg1init}
\end{figure}
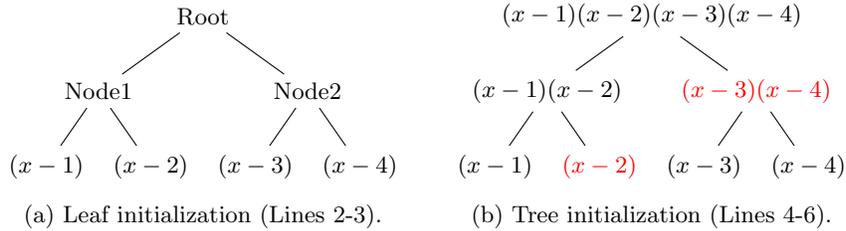

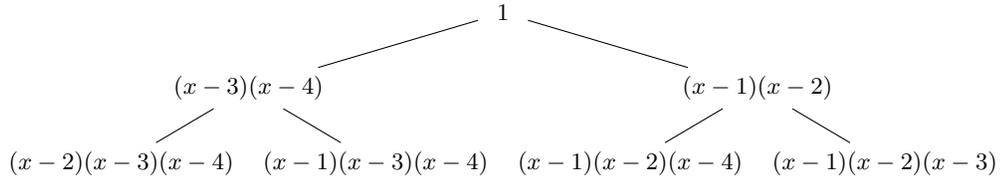
\begin{figure}[ht!]
    \centering
        \begin{tikzpicture}
            \Tree [.$1$
                [ $(x-2)(x-3)(x-4)$ $(x-1)(x-3)(x-4)$ ].$(x-3)(x-4)$
                [ $(x-1)(x-2)(x-4)$ $(x-1)(x-2)(x-3)$ ].$(x-1)(x-2)$
                  ]
        \end{tikzpicture}
    \caption{Algorithm \ref{alg:tree2} computing $T_2$, where the leaves correspond to the one-hot map values of $x$.}
    \label{fig:alg2}
\end{figure}

Algorithm \ref{alg:tree2} is another alternative that requires $2n$ multiplications instead of $n\log{n}$ but has a multiplication depth of $2\log{n}$ instead of $\log{n}$. The algorithm is similar to Algorithm \ref{alg:tree1} except that in Lines 8-10 we group the computation of similar terms. However, because at this step of the algorithm we start from the root and not from the leaves we already used a multiplication depth of $\log{n}$. Figure \ref{fig:alg2} illustrates table $T_2$, which is associated with $T$ from Figure \ref{fig:alg1init}.

 \begin{algorithm}[t!]
    \caption{Small but less shallow}
    \label{alg:tree2}
    \begin{algorithmic}[1]
        \Statex \textbf{Input:} $x$ the input ciphertext, $T$ a balanced binary tree of depth $\log{n}$
        \Statex \textbf{Output:} $o$ a one-hot map that represents $x$.
        \Statex \textbf{Assume:} $\vect{S}$, where $\vect{S}[c]=\prod_{i \in [n], i \neq c}{(c - i)}$
        \Procedure{Opt 1}{}
            \For{$c$ in $[n]$}
                \State $T.\leaf(i) = x-c$
            \EndFor
            \For{$l$ in $\log{n} - 1$ to 0}
                \For{$v$ in $T.\getNodesAtLevel(l)$}
                    \State $v.val = v.\leftn.\val \cdot v.\rightn.\val$
                \EndFor
            \EndFor
            \State Set $T_2=T$ and $T_2.Root.\val = 1$
            \For{$l$ in 1 to $\log{(n)}$-1}\\
               \Comment{$zip$ merges two lists by interleaving their elements.}
                \For{$(v_T, v_{T_2})$ in $zip(T.\getNodesAtLevel(l), T_2.\getNodesAtLevel(l))$}
                    \State $v_{T_2}.\val = (v_{T_2}.\parent.\val) \cdot (v_{T}.\sibling.\val)$
                \EndFor
            \EndFor
            \For{$l$ in $T_2.\leafs()$}
                \State $l.\val = l.\val  \cdot \vect{S}[l]$
            \EndFor
            \State \Return $T_2.\leafs$
        \EndProcedure
    \end{algorithmic}
\end{algorithm}

When running Algorithms \ref{alg:tree1} and \ref{alg:tree2} under encryption the size of $S_c$ may be too small and the size of the product on the intermediate nodes in Line 6 of both algorithms may be too large, which can cause overflows. To this end, we replace Line 8 in Algorithm \ref{alg:tree1} or Line 12 of Algorithm \ref{alg:tree2} with an additional plaintext-ciphertext multiplication per intermediate nodes at Line 6 of both algorithms. The exact plaintext values are stored in a pre-computed ``shadow'' tree that is described in Appendix \ref{sec:sc}. 

\subsection{From Binary representation to one-hot maps}

The trivial way to translate from binary representation to one-hot map representation is to move to a numeric representation by using $a=\sum_{i=0}^{\log{(n)}-1}{\vect{a}[i] \cdot 2^i}$ and then use one of the methods above to generate a one-hot map. Instead, we present another approach reminiscent of the methods we used over \gls{CRT} represented inputs. As before we duplicate the binary values $\vect{a}$ to have arrays of $n$ slots and then return the slot-wise product of these arrays. Only here, the duplication is done using the following trick. For every bit, in slot $x$ we put $\vect{a}[i]$ if $x \pmod{2^i} = 1$ and $1-\vect{a}[i]$ otherwise. In practice, the $\log{n}$ bits plaintext masks $\vect{w}_{i}$ can be pre-computed and multiplied with $\vect{a}[i]$ on the fly using one plaintext-ciphertext (slot-wise) multiplication through the equation
\[
dup(\vect{a}[i]) \cdot \vect{w}_{i} + dup(1-\vect{a}[i])(1-\vect{w}_{i}) = 1 - \vect{w}_{i} + dup(\vect{a}[i]) (2\vect{w}_{i} -1)
\]
We can also achieve the above through selective masking and a rotate-and-sum algorithm.

\subsection{From one-hot representation to binary or CRT representations.}
For completeness, we also provide a transformation from the one-hot representation to binary or CRT representations. The idea is to use predefined fix plaintext masks that represent the required basis. Assume a one-hot map $\vect{o}$ of size $n$. For binary representation we use $n$-element masks $\vect{m}_k$, $k \in [\log{(n)}]$, where $\vect{m}_k[i]= i \pmod{2^{k+1}}$, $i \in [n]$. The conversion is done by computing $\vect{a}[k] = \angles{\vect{m}_k}{\vect{o}}$, where $\vect{a}$ is the output representation. For the CRT case, assume $n=p_1 \cdot p_2 \cdots p_\ell$, then the $\ell$ masks are $\vect{m}_k[i]= i \pmod{p_k}$ and the output is $\vect{a}$, where  $\vect{a}[k] = \angles{\vect{m}_k}{\vect{o}}$.

\subsection{From one-hot representation to CRT maps}

\begin{example}
Let $n=8$ and $a=3$, then the binary representation of $a$ is $\vect{a}=(1,1,0)$ and the duplication maps $\vect{a}^d_{i}$ are:
\begin{align*}
& \vect{a}^d_{0} = (1-\vect{a}[2], \vect{a}[2], 1- \vect{a}[2], \vect{a}[2], 1-\vect{a}[2], \vect{a}[2], 1-\vect{a}[2], \vect{a}[2]) =
(0,1,0,\mathbf{1},0,1,0,1)\\
& \vect{a}^d_{1} = (1- \vect{a}[1], 1- \vect{a}[1],  \vect{a}[1], \vect{a}[1], 1- \vect{a}[1], 1- \vect{a}[1], \vect{a}[1], \vect{a}[1]) =
(0,0,1,\mathbf{1},0,0,1,1)\\
& \vect{a}^d_{2} = (1-\vect{a}[0], 1-\vect{a}[0], 1- \vect{a}[0], 1-\vect{a}[0],
\vect{a}[0], \vect{a}[0], \vect{a}[0],  \vect{a}[0]) = 
(1,1,1,\mathbf{1},0,0,0,0)
\end{align*}
\end{example}

\section{One-hot maps applications}\label{sec:app}
One-hot maps are used by \gls{ML} applications such as decision trees, in different ways. In this section, we review and compare some of these applications. 
\subsection{A single element comparisons}

Comparing integers or fixed point numbers can be costly when performed under \gls{HE}, and specifically CKKS. Some comparison methods include floating-point comparison e.g., in \cite{cheon2020efficient, lee2021minimax} or bit-vector comparisons. Floating-point comparisons often rely on polynomial approximations of the $\Step()$ or $\Sign()$ functions, whereas the accuracy and performance of these methods rely on the degrees of these polynomials. We denote these comparison functions by $\EQ(x,y) \in [1-\beta, 1+\beta] \Leftrightarrow |x - y| < \alpha$, and $\EQ\in [-\beta, \beta]$ otherwise, where $\alpha, \beta > 0$ are parameters that can be made arbitrarily small by increasing the polynomial degree of $\EQ$. Following \cite{cheon2020efficient, lee2021minimax} the degree of $\EQ$ is $degree(\EQ) = O(-\log (\alpha\beta))$.

\paragraph{Bit-vector comparisons.} In contrast, in bit-vector comparisons, the numbers are represented using bit-vectors. For example, the numbers a and b are represented using the $n$-bit vectors $\vect{a},\vect{b} \in \{0,1\}^n$, where $a=\sum_{i=0}^{n-1}{\vect{a}[i]\cdot 2^i}$ and $b=\sum_{i=0}^{n-1}{\vect{b}[i] \cdot 2^i}$. To compare these vectors, one needs to run the circuit $\AND_{i=0}^{n-1}(\XNOR_{i=0}^{n-1}(\vect{a}[i], \vect{b}[i]))$, where for $x,y \in \{0,1\}$, the gates \XNOR (Not Xor) and \AND are implemented using only additions and multiplications with $\XNOR(x,y)=1-(x-y)^2$ and $\AND(x,y)=x \cdot y$. Consequently, for two $n$ bits vectors, the number of multiplications is $2n$ and the multiplication depth is $\log(n)+1$.

\paragraph{Optimization.}
We propose a combination of the approaches above in cases, where the data is already represented in bit-vectors. Specifically, we compute the comparison results by first computing 
$
s = \sum_{i=0}^{n-1}{\XOR(\vect{a}[i], \vect{b}[i])}
$
where a \XOR gate is simulated by $\XOR(x,y)=(x-y)^2$. Then we apply the equality function $1 - \EQ(s, 0)$. Here, $s = 0$ if and only if the numbers are equal. Moreover, because $s \le n$, we can use a dedicated comparison function that only requires $\log(n)$ multiplications as mentioned in \cite{bleach}. The total number of multiplications in our approach is $n + \log(n)$ and the multiplication depth is $\log(n) + 1$. 

We further optimize this method by using the complex plane in CKKS. We first pack every sequential pair of bits as a complex number, where the final complex vectors are $\vect{a^c}[i] = (\vect{a}[2i] + i \vect{a}[2i+1])$ and $\vect{b^c}[i] = (\vect{b}[2i] + i \vect{b}[2i+1])$. These vectors use half the space compared with the previous approach. We now replace the \XOR operations with norm operations, i.e., we first subtract $\vect{d^c} = \vect{a^c} - \vect{b^c}$, and compute
\[
0 \le \sum_{i=0}^{\frac{n}{2}-1}{\vect{d^c}[i] \cdot \overline{\vect{d^c}}[i]} = 
\sum_{i=0}^{\frac{n}{2}-1}{\lVert \vect{d^c}\rVert^2} \underset{\lVert \vect{d^c}\rVert^2 < 2}{\le} 2 \cdot \frac{n}{2} = n
\]
The last inequality follows because every element indicates whether two compared bits are equal (value 0) or unequal (value one of $\pm1, \pm i, \pm1 \pm i$). The rest of the computation is the same. Note that the number of multiplications is now $\frac{n}{2}+\log(n)$ instead of $n+\log(n)$ but we also require $\frac{n}{2}$ conjugate operations.

\paragraph{Using one-hot maps.} Using one-hot maps allows for speeding up the comparison process in cases where the range of inputs can be separated into a not-too-large number of categories. Here, every bit of the one-hot map represents one category. Subsequently, when an application wants to learn whether a number represented using a map represents a specific category, it can simply use the relevant bit from the bit-map.

\begin{example} 
Consider an element $a \in [7]$, specifically, $n=7$ with the one-hot map $\vect{o}_a=(0,0,0,1,0,0,0)$. An application can check whether $n=3$ by either performing $\EQ(3,a)$ or by simply returning $\vect{o}_a[3]$ with $0$ costs.
\end{example} 

\subsection{Greater-equal comparisons}
Another application of one-hot maps is greater, greater-equal, less-than, or, less-equal-than comparisons e.g., $n>10$. Here, we can use a greater-equal (\GE) operator through the polynomial approximation of the $\Step()$ or $\Sign()$ functions from above \cite{cheon2020efficient, lee2021minimax}. When numbers are represented using binary representation, it is possible to efficiently convert them to numbers using $n$ scalar-ciphertext encryptions and additions and then use the \GE operation from above. 
A more efficient method exists using greater maps, which were also used in \cite{crf}.

\begin{definition}
A greater (resp. less-than) map for $n$ categories that represents all categories with indices greater (resp. less-than) from some category index $a$ is an $n$-bits vector $\vect{g}_a \in \{0,1\}^n$ where 
\[
\vect{o}[i]=
\begin{cases}
0 & i \le a~~(resp.~~i \ge a) \\
1 & i > a~~(resp.~~i < a)
\end{cases}
\]
\end{definition}

To perform a greater operation between two values $a$ and $b$ we need (w.l.o.g.) a greater map $\vect{g}_a$ of $a$. In addition, we distinguish between two cases, where $b$ is encrypted or not. If $b$ is unencrypted, we can just return $\vect{g}_a[b]$. Otherwise, either a costly lookup table is required, or we need to represent $b$ as a one-hot map $\vect{o}_b$. Using $\vect{g}_a$ and $\vect{o}_b$ an application can compute 
$
\sum_{i=0}^{n-1}{\AND{(\vect{o}_b[i], \vect{g}_a[i])}}
$
which requires $n$ multiplications and a multiplication depth of $1$.

We can generate a greater map $\vect{g}$ from a one-hot map $\vect{o}$ by going over the bits of $\vect{g}$ from $i=0$ to $i=n-1$ and for every  $i$ compute $\vect{g}[i] = \OR(\vect{o}[i], \vect{g}[i-1])$, where \OR is simulated by $\OR(x,y) = x + y - xy$. Since $\vect{a}$ is given in a one-hot representation a more efficient way is to set $\vect{g}[j] = \sum_{i=0}^{j} \vect{a}[i]$. The cost is $n$ sequential additions that can be done offline. 

\subsection{A range comparison}
The final application is range comparisons, which attempt to find whether a number $m$ is in the range $[a,b]$. This can be done by calling $\AND(\GE(a,n), \GE(n,b))$, or by using a similar algorithm that uses one-hot and greater maps as explained above. 

\section{Packing and tile tensors}\label{sec:pack}

So far we discussed our methods using vectors and operations in vectors. In a real \gls{HE} application these vectors are stored inside ciphertexts, and the operations involve \gls{SIMD} operations on these ciphertexts. 

The way the vectors are stored inside the ciphertexts is called the packing scheme, and it can significantly impact both bandwidth and latency. Consider for example $m$ one-hot maps of size $n$ that require a total of $mn$ elements to pack, and consider ciphertexts of size $s$, such that $m\leq s$ and $n\leq s$, and let's further assume that all three values are close to each other. Two different packing options come to mind: we can store each of the $m$ elements in a single ciphertext containing the entire one-hot vector. Alternately, we can store each number spread across $n$ ciphertexts, where the $i$'th slot in each ciphertext contains the one-hot encoded value for the $i$'th sample.

Recently a new data structure called tile tensor was proposed \cite{aharoni2022complex}. It allows to easily specify these two extreme options and also to easily specify intermediate packing. This is done by writing packing-oblivious code that operates on tensors. The tensors are partitioned to parts (tiles) that are mapped to slots of ciphertexts. Setting a different {\it shape} to the tiles yields a different packing scheme. Consider the matrix of size $[m,n]$ where each of the $m$ rows is the one-hot encoded vector of the $m$'th sample. The tile tensor shape notation $[m/t_1,n/t_2]$ means the matrix is divided into (sub-blocks) tiles of size $[t_1,t_2]$, and each tile is stored in one ciphertext, so that $s=t_1 t_2$.

For example, $[m/1,n/s]$ means the tile's shape is $[1,s]$, so each row of the matrix is stored in a separate ciphertext. Alternately $[m/s,n/1]$ means each column is stored in a separate ciphertext. Other, more complex options are also possible. For example, if $s=1024$, we can choose $[m/16,n/64]$, meaning our matrix is divided into tiles of size $[16,64]$, and each one is stored (flattened) in a ciphertext. To learn more about tile tensor capabilities see \cite{aharoni2022complex}.

Different packings go along well with different algorithms, and the final choice should be made by simulations and measurements with a particular hardware and \gls{HE} backend. Consider for example converting numerical to one-hot representations. We can place one number in a ciphertext, duplicate it across multiple slots, and then it's efficient to run a single comparison in \gls{SIMD} fashion with multiple values. Alternately, we can place multiple numbers in a single ciphertext, where the tree-based approaches we have shown are more efficient.

\input{experiments}

\section{Conclusions}\label{sec:conc}
One-hot maps are a vital tool when considering AI and in particular when running some PPML solutions under HE. We explored some different representations of the (encrypted) input and explain how to move between representations as needed. Our experiment results show the practicality of these translations. Consequently, data scientists have now more tools in their AI-over-HE toolbox that they can efficiently use. In future research, we intend to further explore the trade-offs we get from using different tile tensor capabilities and our algorithms.

\bibliography{main} 
\bibliographystyle{splncs04}

\appendix

\section{Computing the $\vect{S}[c]$ tree}\label{sec:sc}

In Section \ref{sec:trees} we explained why we need to modify Algorithms \ref{alg:tree1} and \ref{alg:tree2} when performed under encryption. Here, provide an algorithm and code to compute the shadow tree that results from the $\vect{S}[c]$ values.

We explain the algorithm by an example. Consider a tree of $n=8$ leaves that represents a one-hot map of 8 elements. The corresponding $\vect{S}[c]$ values are:
\begin{align*}
\vect{S}[0] = \prod_{i=-7}^{-1}{i^{-1}} & & 
\vect{S}[1] = \prod_{\substack{i=-6 \\ i\neq 0}}^{1}{i^{-1}} & &
\vect{S}[2] = \prod_{\substack{i=-5 \\ i\neq 0}}^{2}{i^{-1}} & &
\vect{S}[3] = \prod_{\substack{i=-4 \\ i\neq 0}}^{3}{i^{-1}} \\
\vect{S}[4] = \prod_{\substack{i=-3 \\ i\neq 0}}^{4}{i^{-1}} & &
\vect{S}[5] = \prod_{\substack{i=-2 \\ i\neq 0}}^{5}{i^{-1}} & &
\vect{S}[6] = \prod_{\substack{i=-1 \\ i\neq 0}}^{6}{i^{-1}} & & \vect{S}[7] = \prod_{i=1}^{7}{i^{-1}}  
\end{align*}
It is easy to see the alternate sign pattern of Observation \ref{obs:sc1} and that Observation \ref{obs:sc2} follows from symmetry. We are now ready to describe Algorithm \ref{alg:calcsc}, which generates the $\vect{s}[c]$ shadow tree. The algorithm gets the number of tree levels $\ell$ as input. For every leaf $c$, it computes the list of indices $\vect{S}_0[c]$ in Steps 3-4. Next, it builds an initial tree $T$ by going from the leaves up to the root. At every level $l$, for every node $i$, it computes the intersection $\vect{S}_{l}[2i] \cap S_{l}[2i+1]$ of its two sons and stores the results in $\vect{S}_{l+1}$. In addition, the algorithm stores in the tree nodes the unique elements that are associated with them (Steps 8-9). The idea is to construct a tree that holds the elements of $\vect{S}[c]$ when going through the $T.\leafpath(c)$. 

\begin{observation}\label{obs:sc1}
For $c \in [n]$, when $c$ is odd $\vect{S}[c]<0$ otherwise $\vect{S}[c]>0$.
\end{observation}

\begin{observation}\label{obs:sc2}
For $c \in [n]$, $\vect{S}[c] = - \vect{S}[n-c] $.
\end{observation}

\begin{figure}[h!]
    \centering
    \begin{tikzpicture}[level distance=1.2cm]
        \Tree [.Root
            [ [ $-7$ $1$ ].$\prod\limits_{i=-6}^{-5}{i}$ [ $-5$ $3$ ].$\prod\limits_{i=1}^{2}{i}$ ].$\prod\limits_{i=-4}^{-1}{i}$
            [ [ $-3$ $5$ ].$\prod\limits_{i=-2}^{-1}{i}$ [ $-1$ $7$ ].$\prod\limits_{i=5}^{6}{i}$ ].$\prod\limits_{i=1}^{4}{i}$
              ]
    \end{tikzpicture}
    \caption{An example of the resulted tree $T$ after the first phase of Algorithm \ref{alg:calcsc} (Steps 1-9) when $n=8$.}
    \label{fig:phase1}
\end{figure}

\begin{algorithm}[t!]
    \caption{Calculate the $\vect{S}[c]$ tree}
    \label{alg:calcsc}
    \begin{algorithmic}[1]
        \Statex \textbf{Input:} $\ell$ tree levels.
        \Statex \textbf{Output:} $T$ a balanced tree.
        \Procedure{CalcScTree}{$\ell$}
            \State $n =2 ^\ell$
            \For{$c \in [n]$}
                \State $\vect{S}_0[c] = \{c, c-1,\ldots,1,-1,\ldots c-n+1\}$
            \EndFor

            \For{$l \in [\ell]$}
                \For{$i \in [|\vect{S}_{l}/2|]$}
                    \State $\vect{S}_{l+1}[i] = \vect{S}_{l}[2i] \cap S_{l}[2i+1]$
                    \State $T_l[2i] = \vect{S}_l[2i] - \vect{S}_{l+1}[i]$
                    \State $T_l[2i+1] = \vect{S}_l[2i+1] - \vect{S}_{l+1}[i]$
                \EndFor
            \EndFor

            \For{$l \in [\ell]$}
                \Comment{Swap locations}
                \For{$i \in [|\vect{S}_{l}]/2|]$}
                    \State $T_l[i], T_l[i+1] = T_l[i+1], T_l[i]$
                \EndFor
            \EndFor     

            \For{$l \in (\ell-1)$ to $0$}
                \For{$i \in [T_l]$}
                    \State $T_l[i] = \left(\dfrac{\prod_{T_l[i])}}{\prod{T_{l-1}[2i]} * \prod{T_{l-1}[2i+1]}}\right)^{-1}$
                \EndFor
            \EndFor     

            \For{$i \in [|T_0|]$}
                \State $T_0[i] = \left(\prod{T_0[i]}\right)^{-1}$
            \EndFor

            \State \Return $T$
        \EndProcedure
    \end{algorithmic}
\end{algorithm}

Figure \ref{fig:phase1} shows the tree $T$ for $n=8$ after Step 9. Note the symmetry of the tree that follows Observation \ref{obs:sc2}. Next, Algorithm \ref{alg:tree1} computes the desired product by multiplying the siblings of the nodes on a path to the leaf. Thus, we need to move the relevant $\vect{S}[c]$ elements to be on that path as well. This is done in Steps 10-12. Finally, we convert the lists of values on the tree nodes to integers by computing their product in step 15. We also divide every node by its sons' product to avoid multiplying a single element twice later on once traversing the tree to compute the final product $\vect{S}[c]$ per leaf. The results are presented in Figure \ref{fig:leaf8sc2}. To understand how the shadow $\vect{S}[c]$ is related to $T$, we present also $T$ in Figure \ref{fig:leaf8}. Furthermore, we demonstrate the final Lagrange interpolation value for leaf $3$ by highlighting in red all the product operands that are used by Algorithm \ref{alg:tree1}.

\begin{figure}[ht!]
    \centering
    \begin{tikzpicture}[level distance=1.5cm]
        \Tree [.Root
            [ [ $1^{-1}$ $-7^{-1}$ ].\node[text=red]{$\frac{\prod\limits_{i=1}^{2}{i^{-1}}}{-(1*7)^{-1}}$}; 
              [ \node[text=red]{$3^{-1}$}; $-5^{-1}$ ].$\frac{\prod\limits_{i=5}^{6}{i^{-1}}}{-(3*5)^{-1}}$ ].$\frac{\prod\limits_{i=1}^{4}{i^{-1}}}{\prod\limits_{i \in [1,2,5,6]}{i^{-1}}}$
            [ [ $5^{-1}$ $-3^{-1}$ ].$\frac{\prod\limits_{i=5}^{6}{i^{-1}}}{-(3*5)^{-1}}$
              [ $-1^{-1}$ $7^{-1}$ ].$\frac{\prod\limits_{i=1}^{2}{i^{-1}}}{-(1*7)^{-1}}$ ].\node[text=red]{$\frac{\prod\limits_{i=1}^{4}{i^{-1}}}{\prod\limits_{i \in [1,2, 5,6]}{i^{-1}}}$};
              ]
    \end{tikzpicture}
    \caption{Results of Algorithm \ref{alg:calcsc} when $n=8$, Red nodes show the multiplications required for computing $\vect{o}[3]$ on Algorithm \ref{alg:tree1} Step 11.}
    \label{fig:leaf8sc2}
\end{figure}

\begin{figure}[ht!]
    \centering
    \begin{tikzpicture}
        \Tree [.Root
            [ [ $(x-1)$ $(x-2)$ ].\node[text=red]{$(x-1)(x-2)$}; [ \node[text=red]{$(x-3)$}; $(x-4)$ ].$(x-3)(x-4)$ ].$(x-1)(x-2)(x-3)(x-4)$
            [ [ $(x-5)$ $(x-6)$ ].$(x-5)(x-6)$ [ $(x-7)$ $(x-8)$ ].$(x-7)(x-8)$ ].\node[text=red]{$(x-5)(x-6)(x-7)(x-8)$};
              ]
    \end{tikzpicture}
    \caption{The tree $T$ generated in Algorithms \ref{alg:tree1} and \ref{alg:tree2} at Step 6. Red nodes show the multiplications required for computing $\vect{o}[3]$ on Algorithm \ref{alg:tree1} Step 11.}
    \label{fig:leaf8}
\end{figure}

Table \ref{tab:bounds} shows the minimal and maximal values of a tree of $n$ nodes. We see that the values are not too extreme when $n\le 2^5$.

\begin{table}[ht!]
   \centering
   \caption{Bounds on the positive values in the $S_c$ tree.}
   \label{tab:bounds}
   \begin{tabular}{|c|c|c|c|c|}
    \hline
        $\log(n)$ & minimal value & maximal value & $\log(\min)$ & $\log(\max)$\\
        \hline
        2 & 0.33300  & $1$              & -1.59 & 0 \\
        3 & 0.14300  & $2.5$            & -2.80 & 01.32\\
        4 & 0.06700  & $15.2$         & -3.91 & 03.92\\
        5 & 0.03200  & $292.5$          &  -4.95 & 08.19 \\
        6 & 0.00134  & $110{,}373$     &  -9.54 & 16.75\\
        7 & $1.54e{-}06$ & $1.58e{+}09$ & -19.30 & 33.88 \\
        8 & $2.03e{-}12$ & $3.27e{+}20$  &  -38.00 & 68.15 \\
        \hline
   \end{tabular}
\end{table} 

\newpage
~
\newpage
\section{An example for the hierarchical CRT representation}\label{sec:exhier}

The following example demonstrate the hierarchical CRT representation.
\begin{example} \label{ex:hier}
Let $n=10{,}000$ and $a=5{,}678$, then the hierarchies are
\begin{align*}
& n_1=\ceil{\sqrt{n}}=100 & & n_2=101 & & m_1 = 201\\
& n_{11}=\ceil{\sqrt{100}}=10 & & n_{21}=\ceil{\sqrt{101}}=11\\
& n_{12}=11 & & n_{22}=12   & & m_2 = 44 \\
& n_{111/112/121/122}=[4, 5, 4, 5]  & & n_{211/212/221/222}=[4, 5, 4, 5] & & m_3 = 36
\end{align*} 
To encode $a$ the client computes the 8 residues values
\begin{align*}
& a_1 = a \pmod{n_1} = 78 & &  a_2=a \pmod{n_2} = 22 \\
& a_{11} = a_1 \pmod{n_{11}} = 8 & &  a_{12} = a_1 \pmod{n_{21}} = 0 \\
& a_{12} = a_2 \pmod{n_{12}} = 1 & &  a_{22} = a_2 \pmod{n_{22}} = 10 \\
& a_{111/112/121/122}=[0, 3, 1, 1] & & a_{211/212/221/222} = [0, 0, 2, 0]
\end{align*} 
and their respective maps: 
\begin{align*} 
 & \vect{o}_{a, 1*} = \left((1,0,0,0), (0,0,0, 1, 0), (0,1,0,0), (0,1,0, 0, 0)\right) \\
& \vect{o}_{a, 2*} = \left((1,0,0,0), (1,0,0, 0, 0), (0,0,1,0), (1,0,0, 0, 0)\right)
\end{align*} 
In this example, the number of required slots is $m_3=36$ which is smaller than $r=38$ above.
\end{example}

\end{document}

%% file: macros.tex
\usepackage{makeidx}  
\usepackage{graphicx, amssymb}
\usepackage{color}
\usepackage{placeins}
\usepackage{float}
\usepackage{tabularx,colortbl}
\usepackage[svgnames]{xcolor}
\usepackage{mathtools}
\usepackage{cite}
\usepackage{framed}
\usepackage{amsmath,amsfonts} 
\usepackage{url}
\usepackage[breaklinks]{hyperref}
\usepackage{breakurl} 
\usepackage{comment}
\usepackage{tabularx}
\usepackage{subcaption}
\usepackage{xspace}
\usepackage{tikz-qtree}

\hypersetup{
  colorlinks   = true, 
  urlcolor     = blue, 
  linkcolor    = black, 
  citecolor   = red 
}

\usepackage[nonumberlist,acronyms,nogroupskip]{glossaries}

\newglossary*{not}{Notation}
\usepackage{doi}

\makenoidxglossaries
\setacronymstyle{long-short}

\usepackage{xspace}
\usepackage{commath}
\usepackage[export]{adjustbox}

\usepackage{algorithm}
\usepackage[noend]{algpseudocode}

\usepackage{etoolbox}\AtBeginEnvironment{algorithmic}{\footnotesize}

\usepackage{listings}

\definecolor{codegreen}{rgb}{0,0.6,0}
\definecolor{codegray}{rgb}{0.5,0.5,0.5}
\definecolor{codepurple}{rgb}{0.58,0,0.82}
\definecolor{backcolour}{rgb}{0.95,0.95,0.92}

\definecolor{gray}{gray}{0.5}
\colorlet{commentcolour}{green!50!black}

\colorlet{stringcolour}{red!60!black}
\colorlet{keywordcolour}{magenta!90!black}
\colorlet{exceptioncolour}{yellow!50!red}
\colorlet{commandcolour}{blue!60!black}
\colorlet{numpycolour}{blue!60!green}
\colorlet{literatecolour}{magenta!90!black}
\colorlet{promptcolour}{green!50!black}
\colorlet{specmethodcolour}{violet}

\definecolor{lgrey}{rgb}{0.9,0.9,0.9}
\definecolor{asparagus}{rgb}{0.53, 0.66, 0.42}

\lstdefinelanguage
   [x64]{Assembler}     
   [x86masm]{Assembler} 
   {morekeywords={VMOVDQA64, VPERMPD, VXORPD, VPCLMULQDQ, %
                  VMOVDQU64, VALIGNQ, VAESENC, VAESENCLAST, 
                  VAESDEC, VAESDECLAST, VBROADCASTI64X2, VPXORQ, 
                  VPXOR, VMOVDQU, VPADD, vpsrldq, vpslldq, 
                  vpshufd,vextracti64x4, vextracti128, vpermb,
                  vpblendvb, aesenc, pshufb, zmm0, xmm, rax, rdi, retq, nopw, rcx,
                  .byte,
                  .long, .set},
    morecomment=[l][\color{asparagus}]{\#}} 

\lstdefinestyle{mystyle}{
    backgroundcolor=\color{lgrey},  
      basicstyle=\scriptsize \ttfamily \color{black} \bfseries,
      breakatwhitespace=false,
      breaklines=true,
      captionpos=b,
      commentstyle=\color{asparagus},
      deletekeywords={...},
      escapeinside={\%*}{*)},
      frame=single,
      keywordstyle=\color{purple},
      identifierstyle=\color{black},
      stringstyle=\color{blue},
      language=[x64]Assembler,
      numbers=left,
      numbersep=5pt,
      numberstyle=\scriptsize\color{black},
      rulecolor=\color{black},
      showspaces=false,
      showstringspaces=false,
      showtabs=false,
      stepnumber=1,
      tabsize=5
}
 
\lstdefinestyle{mystyleC}{
    backgroundcolor=\color{lgrey},  
      basicstyle=\scriptsize \ttfamily \color{black} \bfseries,
      breakatwhitespace=false,
      breaklines=true,
      captionpos=b,
      commentstyle=\color{asparagus},
      deletekeywords={...},
      escapeinside={\%*}{*)},
      frame=single,
      keywordstyle=\color{purple},
      identifierstyle=\color{black},
      stringstyle=\color{blue},
      language=C,
      numbers=left,
      numbersep=5pt,
      numberstyle=\scriptsize\color{black},
      rulecolor=\color{black},
      showspaces=false,
      showstringspaces=false,
      showtabs=false,
      stepnumber=1,
      tabsize=5
}

\lstdefinestyle{mystyleP}{
    backgroundcolor=\color{lgrey},  
      basicstyle=\footnotesize \ttfamily \color{black} \bfseries,
      breakatwhitespace=false,
      breaklines=true,
      captionpos=b,
      commentstyle=\color{asparagus},
      deletekeywords={...},
      escapeinside={\%*}{*)},
      frame=single,
      keywordstyle=\color{purple},
      morekeywords={},
      identifierstyle=\color{black},
      stringstyle=\color{blue},
      language=Python,
      numbers=none,
      numbersep=5pt,
      numberstyle=\tiny\color{black},
      rulecolor=\color{black},
      showspaces=false,
      showstringspaces=false,
      showtabs=false,
      stepnumber=1,
      tabsize=5,
      title=\lstname
}

\lstnewenvironment{asmcode}[1][]%
  {\minipage{0.96\linewidth}\medskip 
   \lstset{style=mystyle,frame=single,#1}}
  {\endminipage}

\lstnewenvironment{ccode}[1][]%
  {\minipage{0.96\linewidth}\medskip 
   \lstset{style=mystyleC,frame=single,#1}}
  {\endminipage}

\lstnewenvironment{pythoncode}[1][]%
  {\minipage{0.96\linewidth}\medskip 
   \lstset{style=mystyleP,frame=single,#1}}
  {\endminipage}

\newtheorem{observation}{Observation}

\newcommand{\circler}{\textsuperscript{\textregistered}}

\newcommand{\R}{\mathcal{R}}

\newcommand{\ceil}[1]{\left\lceil #1 \right\rceil}

\newcommand{\angles}[2]{\left\langle #1, #2 \right\rangle}

\newcommand{\Math}[1]{\ensuremath{#1}\xspace}
\newcommand{\AND}{\Math{\operatorname{AND}}}
\newcommand{\XNOR}{\Math{\operatorname{XNOR}}}
\newcommand{\XOR}{\Math{\operatorname{XOR}}}
\newcommand{\OR}{\Math{\operatorname{OR}}}
\newcommand{\EQ}{\Math{\operatorname{Eq}}}
\newcommand{\GE}{\Math{\operatorname{GE}}}
\newcommand{\Sign}{\Math{\operatorname{Sign}}}
\newcommand{\Step}{\Math{\operatorname{Step}}}

\newcommand{\leaf}{\Math{\operatorname{leaf}}}
\newcommand{\leafs}{\Math{\operatorname{leaves}}}
\newcommand{\sibling}{\Math{\operatorname{sib}}}
\newcommand{\leftn}{\Math{\operatorname{l}}}
\newcommand{\rightn}{\Math{\operatorname{r}}}
\newcommand{\val}{\Math{\operatorname{val}}}
\newcommand{\parent}{\Math{\operatorname{parent}}}
\newcommand{\leafpath}{\Math{\operatorname{path}}}
\newcommand{\getNodesAtLevel}{\Math{\operatorname{getNodesAtLevel}}}

\newcommand{\Enc}{\Math{\operatorname{Enc}}}
\newcommand{\Dec}{\Math{\operatorname{Dec}}}
\newcommand{\vect}[1]{\Math{\mathbf{#1}}}
\newcommand{\Cat}{\Math{\vect{c}}}

\renewcommand{\paragraph}[1]{{\medskip\noindent {\bf{#1}}~}}

\usepackage{tikz}
\usetikzlibrary{svg.path}
\definecolor{orcidlogocol}{HTML}{A6CE39}
\tikzset{
  orcidlogo/.pic={
    \fill[orcidlogocol] svg{M256,128c0,70.7-57.3,128-128,128C57.3,256,0,198.7,0,128C0,57.3,57.3,0,128,0C198.7,0,256,57.3,256,128z};
    \fill[white] svg{M86.3,186.2H70.9V79.1h15.4v48.4V186.2z}
                 svg{M108.9,79.1h41.6c39.6,0,57,28.3,57,53.6c0,27.5-21.5,53.6-56.8,53.6h-41.8V79.1z M124.3,172.4h24.5c34.9,0,42.9-26.5,42.9-39.7c0-21.5-13.7-39.7-43.7-39.7h-23.7V172.4z}
                 svg{M88.7,56.8c0,5.5-4.5,10.1-10.1,10.1c-5.6,0-10.1-4.6-10.1-10.1c0-5.6,4.5-10.1,10.1-10.1C84.2,46.7,88.7,51.3,88.7,56.8z};
  }
}
\newcommand{\OrigHeightRecip}{0.00390625}
\newlength{\curXheight}
\DeclareRobustCommand\orcid[1]{%
\texorpdfstring{%
\setlength{\curXheight}{\fontcharht\font`X}%
\href{https://orcid.org/#1}{\XeTeXLinkBox{\mbox{%
\begin{tikzpicture}[yscale=-\OrigHeightRecip*\curXheight,
xscale=\OrigHeightRecip*\curXheight,transform shape]
\pic{orcidlogo};
\end{tikzpicture}%
}}}}{}}

\newcommand{\todo}[1][]{\textcolor{red}}

%% file: experiments.tex
\section{Experiments}\label{sec:exp}

\begin{figure}[ht!]
    \centering
    \includegraphics[width=0.98\textwidth]{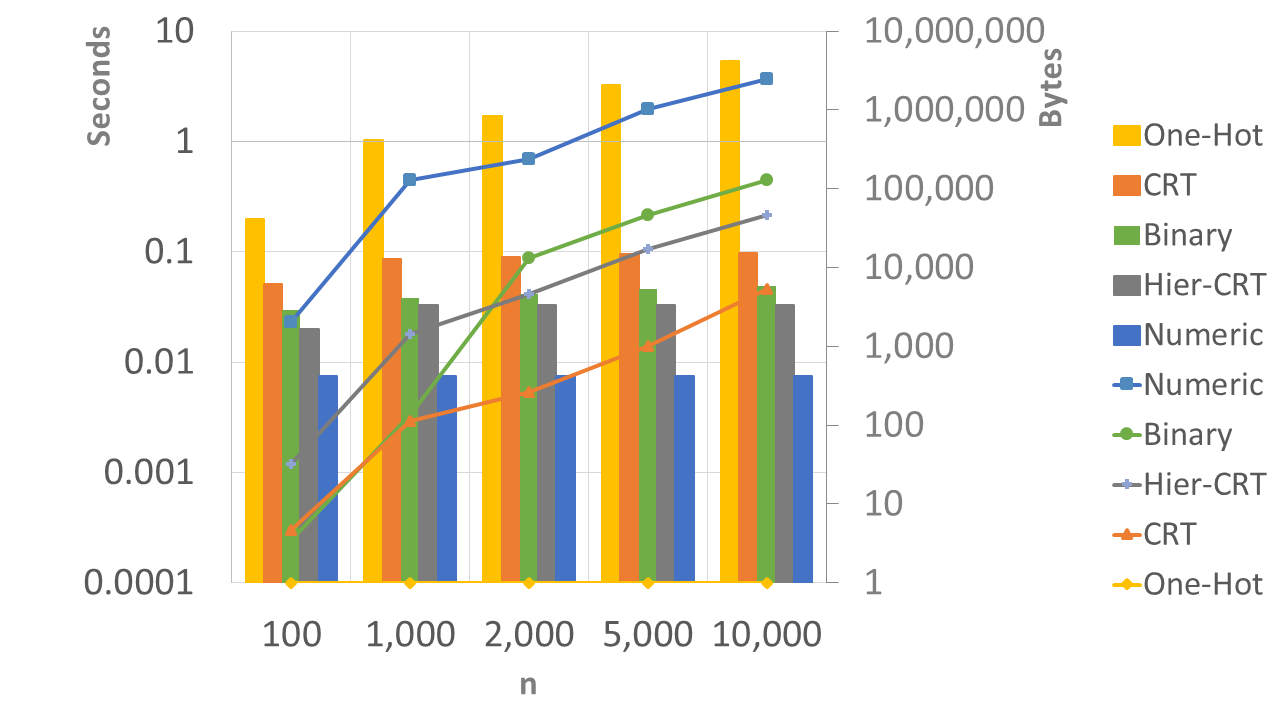}
    \caption{A comparison of the amortized bandwidth and amortized latency per element of generating one-hot maps from different input representations for a batch of $m$ elements and different values of $n$, when setting the tile tensor shape to $[\frac{n}{1}, \frac{m}{s}]$. Scales for both y-axes is logarithmic.}
    \label{fig:onehot}
\end{figure}

For the experiments, we used an Intel\circler Xeon\circler CPU E5-2699 v4 @ 2.20GHz machine with 44 cores (88 threads) and 750GB memory. All the reported results are the average of $10$ runs. We use HElayers~\cite{helayers} version $1.5.2$ and set the underlying HE library to HEaaN\cite{heaanCode} targeting $128$ bits security. 

Figure \ref{fig:onehot} shows our first experiment results, where we measured the amortized bandwidth-latency tradeoff per element in a batch of $m=s$ elements for
generating one-hot maps from different input representations, for different $n$ values. In this experiment we consider five $n$ values from small $n=100$ value to large $n=10{,}000$ value and the five different input representations: a) one-hot map of size $n$; b) numeric representation, the input is a number $a \in [n]$; c) \gls{CRT}, where for example for $n=100$, the input is three one-hot arrays of sizes $3$, $5$, and $7$ ($3\times 5 \times 7 = 105$), and for $n=10{,}000$ the input is five one-hot arrays of sizes $2,5,7,11, 13$ ($2\times 5\times 7\times 11\times 13 = 10{,}010$); 
d) binary, for $n=100$ and $n=10,000$ the client sends $7$ and $14$ bits, respectively; 
d) hierarchical CRT, where for example for $n=100$, the client sends one-hot arrays for 
\[\left((a \bmod{10})\bmod{3},  (a \bmod{10})\bmod{4}, (a \bmod{11})\bmod{3},  (a \bmod{11})\bmod{4} \right)
\]
of sizes $3,4,3,4$, respectively, and for the $n=10{,}000$ case, the client sends the maps of sizes $3,4,3,4,3,4,3,4$ from which a one-hot array for $a \in [10{,}000]$ can be computed.

\paragraph{HE parameters.} We used a bootstrappable context with ciphertexts with $s=2^{15}$ slots, a multiplication depth of $12$, fractional part precision of $42$, and an integer part precision of $18$. In all cases, we used a batch $m$ of $s$ samples that are packed together in one ciphertext.
For the $n=100$ case, we set \EQ to be a polynomial of degree $7^4$ (a composition of $4$ polynomials of degree $7$) with a multiplication depth of $12$, the resulting maximum error was 0.02. For the $n=10{,}000$ case, we used a larger polynomial of degree $7^7$, with a multiplication depth of 21. The resulting maximum error was 0.04. The error was virtually 0 in the CRT and Hier-CRT cases.

As expected, Figure \ref{fig:onehot} shows that the extreme case is when the client sends the entire one-hot array and the server does nothing (yellow graph). The other extreme case is when the client transmits a single number (numeric representation) and the server does a full transformation (blue graph). In all cases, we measured the cost to generate a one-hot map under encryption. We did not implement or measure the applications that use the one-hot maps, such as decision trees, because that would only distract and obscure the measurements we are interested in, namely how efficient is computing one-hot maps. We refer the reader to e.g., \cite{crf} to learn about the performance of an encrypted tree based solution. 
\paragraph{Numeric to one-hot} We also tested the performance and accuracy of Algorithms \ref{alg:tree1} and \ref{alg:tree2} for converting a number to its one-hot representation when using tile tensors of shape $[n/1,m/s]$, where $n$ is the one hot size and $m$ is the number of samples that run in parallel, in our case, we set $m=s$. Table \ref{tab:numtoonehot} summarizes our experiment results. Here, we used ciphertexts with $2^{14}$ slots, a multiplication depth of $17$, fractional part precision of $44$, and an integer part precision of $16$. 

Our experiments showed that using  Algorithms \ref{alg:tree1} and \ref{alg:tree2} directly was only possible for small $n \le 2^3$ values because the tree values get too large and cause overflows. However, when combining them with the shadow tree method presented in Appendix \ref{sec:sc}, we were able to get good results even for larger $n$ of size $2^5=32$. Thus, our algorithms present another tradeoff between latency and accuracy that a user can exploit. 

\begin{table}[t!]
    \centering
    \caption{Performance and accuracy results of Algorithms \ref{alg:tree1}, \ref{alg:tree2} and using \EQ for different $n$ values. The reported precision is the maximal measured absolute distance from the expected result.}
    \label{tab:numtoonehot}
    \begin{tabular}{|c|c|c|c|c|c|c|}
        \hline
        $n$ & \multicolumn{2}{c|}{Algorithm \ref{alg:tree1}} & \multicolumn{2}{c|}{Algorithm \ref{alg:tree2}} & \multicolumn{2}{c|}{Using \EQ} \\
        & Latency (sec) & Precision & Latency (sec) & Precision & Latency (sec) & Precision \\
        \hline
        $2^2$ & 0.392 & 8.7e-9 & 0.355 & 9.1e-9 & 0.807 & 3.8e-8 \\
        $2^3$ & 1.085 & 1.7e-07 & 0.662 & 2.8e-7 & 1.499 & 2.9e-8 \\
        $2^4$ & 3.422 & 6.2e-06 & 1.917 & 3.1e-5 & 4.568 & 1.1e-7 \\
        $2^5$ & 9.138 & 0.24 & 6.096 & 0.3 & 11.080 & 5.2e-7 \\
        $2^6$ & 38.684 & $\infty$ & 25.498 & $\infty$ & 50.727 & 2.3e-6 \\
        \hline
    \end{tabular}
\end{table}